\documentclass[journal=jpcbfk,manuscript=article]{achemso}
\usepackage{longtable}
\usepackage{multirow}
\usepackage{amsmath}
\usepackage[usenames,dvipsnames]{color}
\usepackage{soul}
\usepackage{threeparttable}
\usepackage{xr}
\usepackage{graphicx}
\usepackage{amsmath,amssymb}
\usepackage{caption}
\usepackage{color}
\usepackage{xcolor}
\usepackage{dcolumn}
\usepackage{bm}
\usepackage{float}
\usepackage{subcaption}
\usepackage{textcomp}
\usepackage{url}
\usepackage[normalem]{ulem}

\usepackage{microtype}

\newcommand{\mm}[1] {{\color{red} #1}}   %MM
\usepackage[unicode=true,
            bookmarks=true,
            bookmarksnumbered=true,
            bookmarksopen=true,
            bookmarksopenlevel=2,
            breaklinks=false,
            pdfborder={0 0 1},
            backref=false,
            colorlinks=true,
            hidelinks
            ]{hyperref}

\makeatletter

\author{Kai T\"opfer} \affiliation[University of Basel]{Department of
  Chemistry, University of Basel, Klingelbergstrasse 80, CH-4056
  Basel, Switzerland.}

\author{Jingchun Wang, Shimoni Patel} \affiliation[University of Basel]{Department of
  Chemistry, University of Basel, Klingelbergstrasse 80, CH-4056
  Basel, Switzerland.}

\author{Markus Meuwly} \affiliation[University of Basel]{Department of
  Chemistry, University of Basel, Klingelbergstrasse 80, CH-4056
  Basel, Switzerland.}
\email{m.meuwly@unibas.ch}

\title{Structure and Dynamics of Deep Eutectic Systems from
  Cluster-Optimized Energy Functions}

\date{\today}

\begin{document}
\date{\today}

\begin{abstract}
Generating energy functions for heterogeneous systems suitable for
quantitative and predictive atomistic simulations is a challenging
undertaking. The present work combines a cluster-based approach with
electronic structure calculations at the density functional theory
level and machine learning-based energy functions for a spectroscopic
reporter for eutectic mixtures consisting of water, acetamide and
KSCN. Two water models are considered: TIP3P which is consistent with
the CGenFF energy function and TIP4P which - as a water model - is
superior to TIP4P. Both fitted models, {\bf M2$^{\rm TIP3P}$} and {\bf
  M2$^{\rm TIP4P}$}, yield favourable thermodynamic, structural,
spectroscopic and transport properties from extensive molecular
dynamics simulations. In particular, the slow and fast decay times
from 2-dimensional infrared spectroscopy and the viscosity for
water-rich mixtures are described realistically and consistent with
experiments. On the other hand, including the co-solvent (acetamide)
in the present case is expected to further improve the computed
viscosity for low-water content. It is concluded that such a
cluster-based approach is a promising and generalizable route for
routine parametrization of heterogeneous, electrostatically dominated
systems.
\end{abstract}

\section{Introduction}
Atomistic simulations are a powerful approach to investigate the
energetics, structural dynamics, and spectroscopy of heterogeneous
systems in the condensed phase. This has, {\it inter alia}, been
demonstrated for hydrated proteins, ionic liquids, or deep eutectic
mixtures.\cite{dodson:2008,klepeis:2009,liu:2012,habasaki:2008,MM.eutectic:2022,gould:2025}
Such studies can ideally complement experimental efforts and provide
molecular-level characterization and
interpretation.\cite{MM.mb:2016,MM.hb:2018} From a computational
viewpoint the main challenges are a) the accuracy of the energy
function to carry out atomistic simulations and b) the time scale on
which such simulations can be run. Conversely, from an experimental
and measurement perspective, one of the great challenges is the fact
that usually the ``full system'' is probed. In other words, for
heterogeneous systems it is difficult to arrive at a molecular-level
structural interpretation for a specific part of the system from
measurements that report on the entire system.\cite{MM.insulin:2020}
Also, it is demanding to cover multiple time scales ranging from the
femtosecond to the second time scale in a single measurement, although
recent progress has been made here.\cite{hamm:2023}\\

\noindent
The quality of atomistic simulations is directly linked to the
accuracy with which the inter- and intramolecular interactions are
described. Ideally, the classical (Newton) or quantum mechanical
(Schr\"odinger) time evolution equations to follow the nuclear
dynamics would be solved by using energies and forces from high-level
quantum chemistry calculations with large basis sets. However, this is
usually not feasible but for the smallest systems (molecules with
$\sim 5$ heavy atoms) and on short time scales (tens to hundreds of
picoseconds). Machine learning-based energy functions have improved
this situation considerably, in particular if more specialized
techniques such as transfer learning (TL) are
employed.\cite{MM.tl:2021,MM.tl:2023,MM.tl:2024,MM.tl:2025}
Nevertheless, routine application of ML-based energy functions to
heterogeneous condensed-phase systems is still not routine. Instead,
targeted improvements of empirical energy functions remain an
attractive alternative as they combine the robustness of a
coordinate-dependent functional dependence with the flexibility of a
parametrized model that can be adapted to either quantum chemical
reference data, experimental observables, or both. Still, one of the
challenges for more physics-based models is to develop energy
functions that retain the precision of the quantum mechanical methods
they are often based on.\\

\noindent
Traditionally, empirical energy functions use harmonic springs to
represent bonds and valence angles, and periodic functions for
dihedrals. For the non-bonded interactions it is common to employ
atom-centered point charges and a Lennard-Jones (LJ) representation
for van-der-Waals (vdW) interactions.\cite{mackerell2004} For
applications to vibrational spectroscopy it is necessary to go beyond
the harmonic approximation and to include effects of mechanical
anharmonicity and coupling between different internal degrees of
freedom. Such improvements can, e.g., be achieved through the use of
machine learning-based
approaches.\cite{MM.n3:2019,MM.jcp:2020,MM.rkhs:2020,nandi:2019,li:2014}\\

\noindent
Similarly, the electrostatic model can be improved by going beyond the
first-order treatment based on atom-centered point charges to better
describing anisotropic contributions to the charge
density.\cite{Stone2013} Including higher-order atomic multipoles
improves the accuracy but at the expense of increased computational
cost and implementation
complexity.\cite{Handley2009,MM.mtp:2013,Devereux2014,bereau:2016}
Accounting for polarizability is another contribution that has been
included in empirical force fields and shows promise for further
improvements of the computational models.\cite{ren:2019} From an
empirical force field perspective the vdW interactions are often
represented as Lennard-Jones terms with {\it ad hoc}
(Lorentz-Berthelot) combination rules. Alternative and potentially
improved representations are the buffered 14-7
parametrization\cite{halgren:1992} and/or modified combination
rules\cite{mason:1988,millie:2001}\\

\noindent
Deep eutectic solvents (DESs) are multicomponent mixtures consisting
of molecular species acting as hydrogen bond acceptors and hydrogen
bond donors at particular molar
ratios.\cite{abbott2003DES,marcus2019trends,martins2019defdes} In DESs
the melting point of the mixture is lower than that of the individual
components.\cite{smith:2014} They also remain in the liquid phase over
a wider temperature range.\cite{smith:2014} If the mixtures contain
ions, the intermolecular interactions involve pronounced electrostatic
contributions which is also - in part - due to crowding. The
particular mixture considered here consists of water, acetamide and
KSCN which is present as solvated K$^+$ and SCN$^-$ (thiocyanate)
ions.\cite{isaac:1988,kalita:1998} Acetamide forms low-temperature
eutectics with a wide range of inorganic salts and the resulting
non-aqueous solvents have a high ionicity. Such mixtures have also
been recognized as excellent solvents and molten acetamide is known to
dissolve inorganic and organic compounds. The SCN$^-$ anion is a
suitable spectroscopic probe because the CN-stretch vibration absorbs
in an otherwise empty region of the infrared region. Advantage of this
has been recently taken to probe the effect of water addition to
urea/choline chloride and in acetamide/water
mixtures.\cite{sakpal:2021,MM.eutectic:2022}\\

\noindent
Computational approaches for deep eutectic solvents have been recently
reviewed.\cite{velez:2022} Previous efforts to parametrize empirical
energy functions for deep eutectic used a range of protocols and
approaches\cite{ferreira:2016,padua:2022,
  maglia:2021,doherty:2018,jeong:2021,garcia:2015,zhang:2022,velez:2022}
In almost all the cases, the starting point was a conventional energy
function such as the General Amber Force Field (GAFF),
\cite{wang:2004} but adapted for particular
applications\cite{perkins:2014} In a next step, particular parameters
were and to readjusted\cite{ferreira:2016,zhang:2022} which included
scaling of partial charges\cite{garcia:2015} and/or scaling of
van-der-Waals parameters to reproduce observed properties such as
diffusivities, viscosities or the densities.\cite{zhang:2022}
Alternatively, models were also developed based on symmetry adapted
perturbation theory (SAPT)\cite{jeong:2021} and to refine them by
comparing with first principles MD simulations. More recent
work\cite{MM.eutectic:2024} focused on using cluster systems extracted
from molecular dynamics (MD) simulations and computing total
interaction energies based on electronic structure
calculations.\cite{MM.ff:2024} Using reference data from density
functional theory calculations, specific force field parameters were
adjusted to best reproduce the reference data. This
protocol,\cite{MM.eutectic:2024} referred to as {\bf M2} in previous
and the present work, can be amended by comparison with available and
reliable measured properties of the system but {\it a priori} no
experimentally measured data is required.\\

\noindent
The present work aims at parametrizing atomistic force fields using
state-of-the art methods by combining machine learning-based
approaches for bonded and nonbonded terms, refinement of the
Lenard-Jones interactions with respect to thermodynamic data and
validation on structural, spectroscopic and thermodynamic
measurements. First, the methods are presented, followed by the
reparametrization and validation of the energy functions. Next,
extended MD simulations are analyzed with respect to pair distribution
functions, frequency fluctuation correlation functions from experiment
and simulations are compared, and the viscosities of different
mixtures are determined. Finally, conclusions are drawn.\\

\section{Computational Methods}

\subsection{Simulation Setup}
Molecular dynamics simulations were carried out using the CHARMM
program\cite{Charmm-Brooks-2009} with provisions for electrostatics
based on the flexible minimal distributed charge
model\cite{MM.fmdcm:2022} (fMDCM) and bonded interactions described by
a reproducing kernel Hilbert space
(RKHS).\cite{MM.charmm:2024,MM.rkhs:2017} The molar composition of the
mixtures was changed by varying the number of water and acetamide
molecules while keeping constant the total concentration of K$^+$ and
SCN$^{-}$, see Table \ref{sitab:composition}. The cutoff for nonbonded
interactions was 14~\AA\/ and electrostatic interactions were treated
using the Particle Mesh Ewald algorithm.\cite{Darden1993} and bonds
involving hydrogen atoms were constrained using the SHAKE
algorithm.\cite{shake77} In total, 5 independent random initial
configurations for each of the 9 system compositions were set up using
PACKMOL.\cite{martinez:2009} After 100\,ps of heating and
equilibration simulation, respectively, {\it NpT} production
simulations were run at 300\,K and normal pressure (1\,atm) were
performed for 5\,ns with a time step of 1\,fs using the leap-frog
integrator and a Hoover thermostat within the extended system constant
pressure and temperature algorithm as implemented in
CHARMM.\cite{Hoover1985,Brooks1995} The mass of the pressure piston
and piston collision frequency were $406$\,u and 5\,ps$^{-1}$,
respectively, and the mass of the thermal piston was
$4060$\,kcal/mol\,ps$^2$.  For each system composition a total of
$25$\,ns was sampled.\\

\subsection{Inter- and Intramolecular Interactions}
The representation and fitting of the intra- and inter-molecular
contributions to the total energy function were described in previous
work.\cite{MM.eutectic:2024} Here, a brief summary is given. The total
energy function for the heterogeneous mixture (water, acetamide,
K$^+$, SCN$^{-}$) was described by a combination of the all-atom
CGenFF force field\cite{cgenff} for acetamide, the TIP3P water
model\cite{TIP3P-Jorgensen-1983} to be used together with CGenFF, and
literature LJ parameters for the potassium cation K$^+$ with an
assigned atom charge of $+1.0$.\cite{bian:2013} For the thiocyanate
anion (SCN$^-$) the bonding potential was a reproducing kernel Hilbert
space (RKHS)\cite{rabitz:1996,MM.rkhs:2017} representation based on
{\it ab initio} data at the PNO-LCCSD(T)-F12/aug-cc-pVTZ-F12 level of
theory. The SCN$^-$ electrostatics were based on the fMDCM
model\cite{MM.fmdcm:2022} fitted to the electrostatic potential (ESP)
calculated at the M06-2X/aug-cc-pVTZ level of theory. For the fMDCM
model, 8 point charges were distributed around the SCN$^-$ atoms with
positions within the local axis frame determined by 3rd order
polynomial functions $f(x)$ with $x=1-\cos^2 \theta$ and the SCN$^-$
bond angle $\theta$.  The 96 parameters of the 24 polynomial functions
- 4 parameters per polynomial for each Cartesian coordinate (3) and
distributed charge (8) - were optimized to best reproduce the
reference ESPs for different SCN$^-$ conformations. \\

\begin{figure}
  \centering
  \includegraphics[width=0.50\textwidth]{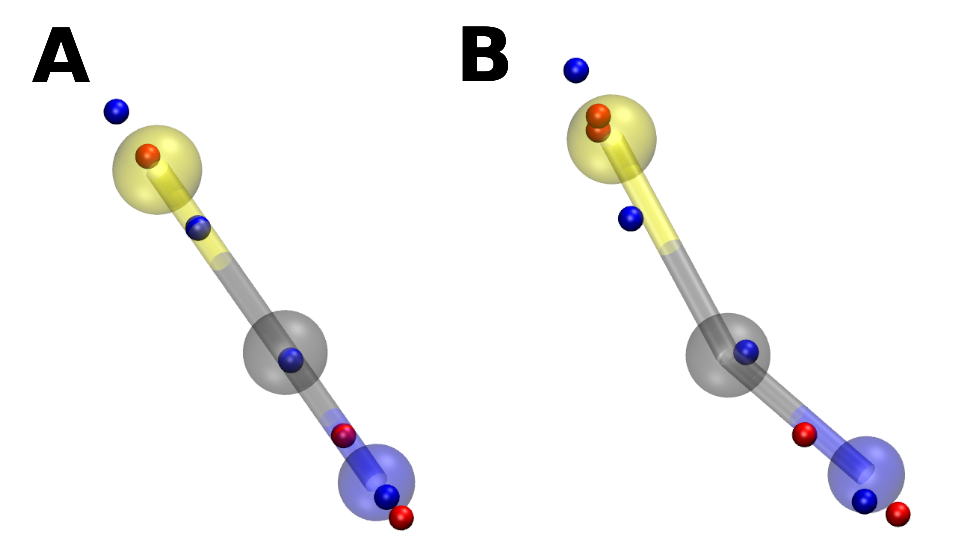}
\caption{Distributed charge positions (red spheres as negative charges
  and blue spheres positive charges) predicted by the fMDCM model to
  represent the ESP of SCN$^-$ (transparent spheres) for the (A)
  linear equilibrium conformation and (B) at a valence angle $\theta =
  160^\circ$.}
\label{fig:fmdcm}
\end{figure}

\noindent
The present work focuses primarily on model {\bf M2} which is a
combination of fMDCM and optimized LJ parameters of SCN$^-$ to best
reproduce {\it ab initio} interaction energies between SCN$^-$ with
cluster shells of different sizes and
composition.\cite{MM.eutectic:2024} For this, the LJ parameters
$\epsilon$ and $r_\mathrm{min}$ of SCN$^-$ were adjusted to best match
counterpoise-corrected\cite{boys:1970} interaction energies from
electronic structure calculations. The systems considered included one
SCN$^-$ anion surrounded by (i) 16 water molecules; (ii) 14 water
molecules with one K$^+$ ion, (iii) 14 water molecules with one
additional SCN$^-$; (iv) 12 water molecules with both, one additional
SCN$^-$ and K$^+$ ion. For each type of system (i) to (iv) 50
independent conformations were extracted randomly from previous MD
simulations.\cite{MM.eutectic:2022} Reference interaction energies
were then determined at the M06-2X/aug-cc-pVTZ level of theory using
the Gaussian program.\cite{gaussian16} Higher levels of quantum
chemical theory, such as coupled cluster energies, are not feasible
due to the unfavourable scaling of the computations with system and
basis set size.\\

\noindent
In addition to the TIP3P model, an independent parametrization was
carried out by using the TIP4P model because the TIP3P model has known
deficiencies.\cite{kadaoluwa:2021} However, strictly speaking, the
TIP4P model is not fully compatible with CGenFF, see above. For the
reference quantum chemistry calculations, SCN$^-$ centered cluster
conformations were extracted from previous MD simulations with TIP4P
water model,\cite{MM.eutectic:2022} but using the previous optimized
LJ parameter for SCN$^-$ in TIP3P water containing clusters.  As a
consequence to changed local densities, the cluster composition (iii)
is adjusted to 13 water molecules with one additional SCN$^-$. Still,
for each type of system (i) to (iv) 50 independent conformations were
extracted randomly and counterpoise-corrected interaction energies
from electronic structure calculations between SCN$^-$ and the cluster
shell were computed. The LJ parameters of SCN$^-$ were optimized to
best match these interaction energies.\\

\noindent
In the past, the TIP3P water model has been found to have various
limitations when comparing experimentally measured quantities with
those from simulations.\cite{kadaoluwa:2021} It should, however, be
kept in mind that the CGenFF energy function was parametrized together
with the TIP3P water model. After fitting the LJ parameters to the
cluster energies from electronic structure calculations it is found
that the two water models yield similarly accurate representations of
the intermolecular interactions. For the remainder of the present
work, simulation results from parametrizations following method {\bf
  M2} - fitting of LJ-parameters based on quantum chemistry for
cluster models - using water models TIP3P and TIP4P (models {\bf
  M2$^{\rm TIP3P}$} and {\bf M2$^{\rm TIP4P}$}) are presented and
discussed.\\
    
\subsection{Analysis}
The hydration free energy $\Delta G_{\mathrm{hyd}}$ for the SCN$^-$
anion in water solvent was computed from thermodynamic
integration.\cite{MM.cn:2013} One SCN$^-$ anion was sampled in the gas
phase and in pure water. The condensed-phase simulations were carried
out in the \emph{NpT} ensemble with 997 water molecules (cubic box
size $\sim 30^3$\,\AA$^3$).\cite{straatsma:1988,MM.cn:2013} The
coupling parameter $\lambda \in (0, 1)$ included 24 evenly spaced
values for the electrostatic and vdW interactions,
respectively. Initial conditions for these simulation were taken from
an unbiased simulation, equilibrated for 50\,ps with the respective
coupling parameter $\lambda$ and run for another 150\,ps for
statistical sampling.  The hydration free energy was then accumulated
from
\begin{equation}
     \Delta G_{\mathrm{hyd}} = \sum_\lambda [ (
       H_\mathrm{solv}^\mathrm{elec}(\lambda) -
       H_\mathrm{gas}^\mathrm{elec}(\lambda) ) + (
       H_\mathrm{solv}^\mathrm{vdW}(\lambda) -
       H_\mathrm{gas}^\mathrm{vdW}(\lambda) ) ] \Delta \lambda
\end{equation}
For a triatomic such as SCN$^-$,
$H_\mathrm{gas}^\mathrm{elec}(\lambda) =
H_\mathrm{gas}^\mathrm{vdW}(\lambda) = 0$ due to the 1-2 and 1-3
nonbonded interaction exclusion.\cite{mackerell:2004} Therefore, only
$H_\mathrm{solv}^\mathrm{elec}(\lambda)$ and
$H_\mathrm{solv}^\mathrm{vdW}(\lambda)$ needed to be accumulated.\\

\noindent
The density of aqueous KSCN solution was determined from 500\,ps
simulations in the $NpT$ ensemble (100\% water in Table
\ref{sitab:composition}) with a KSCN molality of
$b(\mathrm{KSCN})=3.821$\,mol/kg. The equilibrium simulation box
volume was computed as the average box volume from the last 100\,ps of
the simulations (400--500\,ps) during the production run.  Convergence
within the reported precision was checked by comparing with the
average taken from the results of the full production run of
500\,ps.\\

\noindent
For the frequency fluctuation correlation function (FFCF), the
frequency trajectories $\omega_i(t)$ for each oscillator $i$ were
determined from an instantaneous normal mode (INM) analysis of the
CN-vibrational frequencies $\omega_i$.\cite{stratt:1994} All SCN$^-$
ions were analyzed on snapshots separated by $100$\,fs along the first
2\,ns of each production simulation (10\,ns in total for each system
composition).  For the INM analysis the structure of each SCN$^-$ ion
was optimized whereby the positions of all remaining atoms in the
system were frozen. This was followed by a normal mode analysis using
the same force field that was employed for the MD
simulations. \mm{Previously, such an approach has been validated for
N$_3^-$ in solution by comparing with rigorous quantum bound state
calculations.\cite{MM.n3:2019}}\\

\noindent
From the frequency trajectory $\omega_i(t)$ for each oscillator the
FFCF $\delta \omega_i(t) = \omega_(t) - < \omega_i >$ was determined
which contains information on relaxation time scales corresponding to
the solvent dynamics around the solute. The FFCFs were fit to an
empirical expression\cite{kozinski:2007,MM.cn:2013}
\begin{equation}
  \langle \delta \omega(t) \delta \omega(0) \rangle = a_{1}
  e^{-t/\tau_{1}} + a_{2} e^{-t/\tau_{2}} + \Delta_0^2
\label{eq:ffcffit}
\end{equation}
using an automated curve fitting function
(\texttt{scipy.optimize.curve\_fit}) from the SciPy library using the
default trust region reflective algorithm.\cite{2020SciPy-NMeth} Here,
$a_{i}$, $\tau_{i}$, $\gamma$ and $\Delta_0^2$ are the amplitudes,
decay time scales, phase and asymptotic value of the FFCF.\\

\noindent
The viscosities $\eta$ for varying acetamide/water ratios were
determined from the stress tensor $\mathbf{P}(t)$ according to $\eta =
\frac{V}{6 k_B T} \sum_{\alpha \le \beta} \int_{0}^{\infty} \langle
\bar{P}_{\alpha\beta}(0) \bar{P}_{\alpha\beta}(t) \rangle \mathrm{d}t$
$(\alpha, \beta = x, y, z)$ where $\bar{P}_{\alpha\beta}$ are the
upper triangular elements of the modified stress tensor
$\mathbf{\bar{P}}(t)$.\cite{visc_zhang:2020,zhang:2022} Due to the
strong intermolecular interactions and high viscosities, converging
$\eta$ can be rather demanding and is not attempted here. Rather, 5
independent $NVT$ simulations of 5\,ns each were carried out for each
composition and the results were averaged to obtain illustrative
results for an average $< \eta >$ and a fluctuation around it.\\

\section{Results}

\subsection{Validation of the Energy Function}
\begin{figure}
  \centering
  \includegraphics[width=0.90\textwidth]{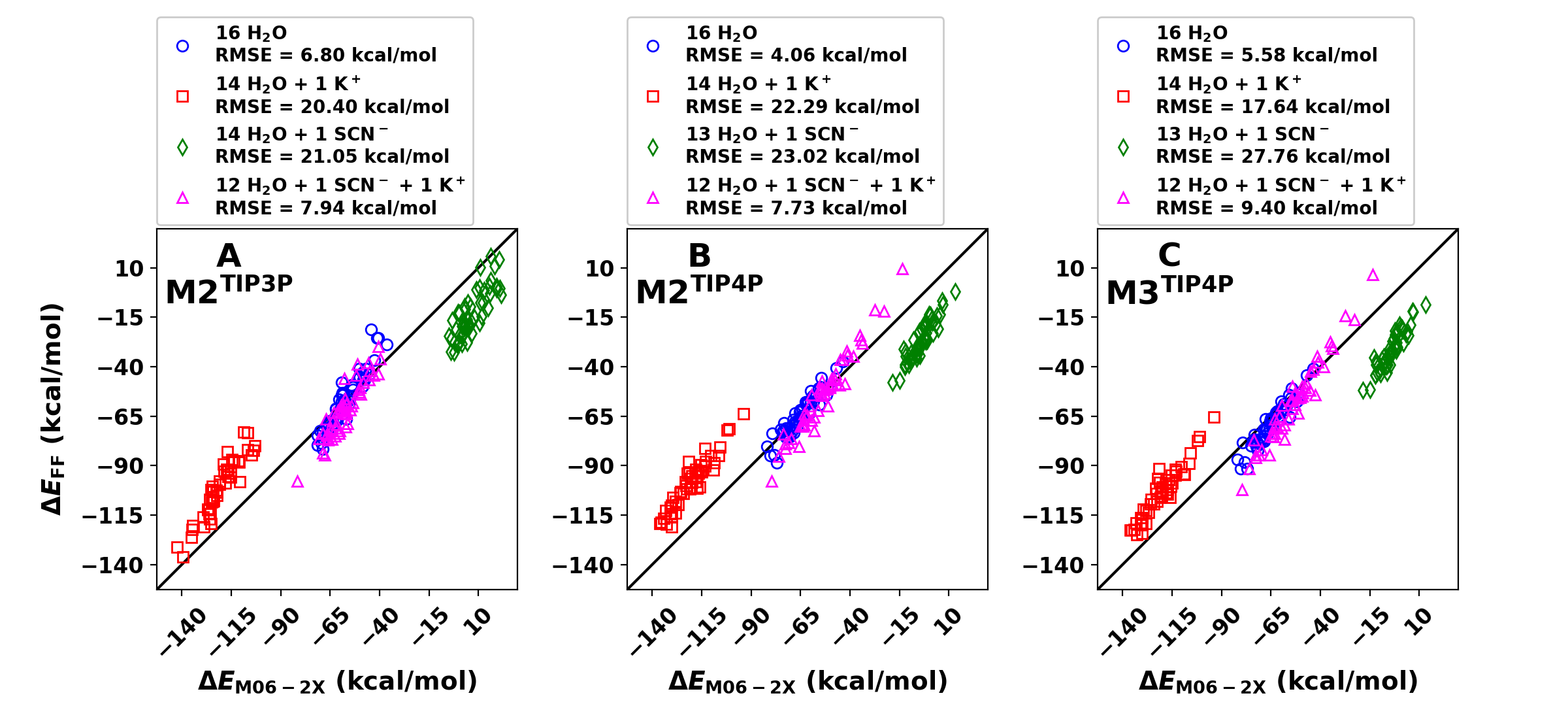}
\caption{Correlation between interaction energies from reference {\it
    ab initio} calculations and those based on the fitted model for
  KSCN using the (A) TIP3P and (B) TIP4P water models. Random
  snapshots were taken from equilibrium simulations,
  respectively. Panel C shows the correlation of setup (B) {\bf
    M2$^{\rm TIP4P}$} but with scaled LJ parameter $r_\mathrm{min}$ by
  a factor of $0.96$, see Figure \ref{fig:rho_hfe}.  The systems
  considered always consist of one SCN$^-$ anion surrounded by shells
  as indicated in the legend with corresponding RMSE. The overall
  RMSEs are $15.56$\,kcal/mol, $16.61$\,kcal/mol, and
  $17.33$\,kcal/mol in panels A to C, respectively.}
\label{fig:models}
\end{figure}

\noindent
To set the stage, the performance of the two optimized energy
functions is compared in Figure \ref{fig:models}. The clusters used in
the parametrization consist of SCN$^-$(H$_2$O)$_{16}$ (blue),
SCN$^-$K$^+$(H$_2$O)$_{14}$ (red), (SCN$^-$)$_2$(H$_2$O)$_{14}$ or
(SCN$^-$)$_2$(H$_2$O)$_{13}$ (green), and
(SCN$^-$)$_2$K$^+$(H$_2$O)$_{12}$ (magenta). For the {\bf M2$^{\rm
    TIP3P}$} the RMSE of 15.56\,kcal/mol compares with 16.61\,cal/mol
for {\bf M2$^{\rm TIP4P}$}. Given the overall energy range of about
150 kcal/mol covered by the data set and the small number of
adjustable parameters (6 LJ parameters), a 10\% difference is
acceptable. In addition, for each of the subsets (i) to (iv) the
correlation between reference and model is clearly established. Both
parametrizations underestimate the attractive interaction energy
between the center SCN$^-$ with K$^+$ in the cluster shell, but
overestimate the interaction energy between the center SCN$^-$ and a
second SCN$^-$ in the cluster shell. The equivalence of the two
parametrizations is even more notable as the cluster geometries were
generated from two entirely independent simulations, one carried out
using the TIP3P water model and the other one using TIP4P. Even the
RMSE for the four subgroups are within 2 kcal/mol of each other.\\

\begin{figure}
  \centering
  \includegraphics[width=0.75\textwidth]{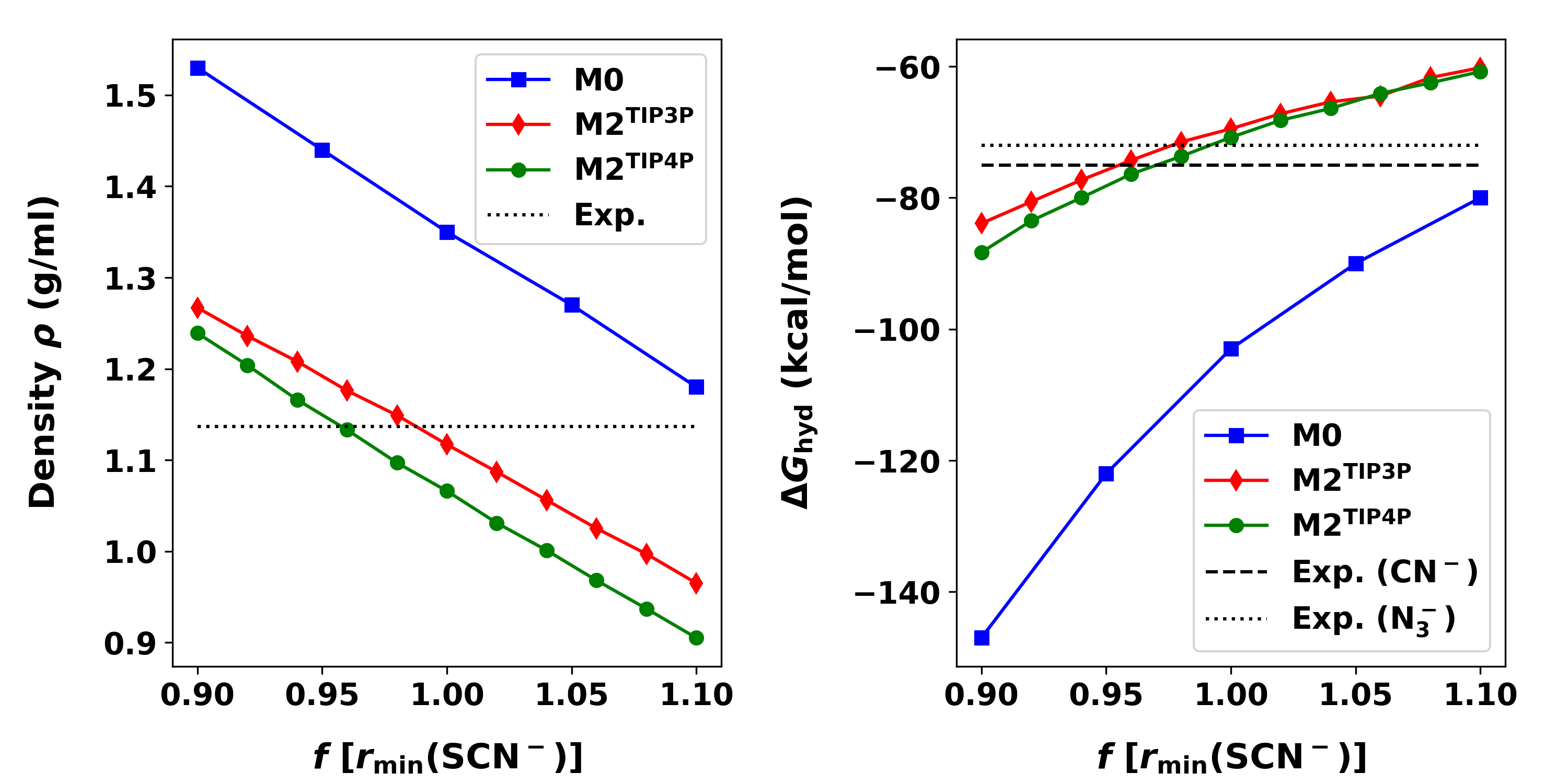}
\caption{Left panel: Computed density of a $3.8$\,mol/kg KSCN solution
  in water at different scaled $r_\mathrm{min}$ values using models
  {\bf M0} (scaled LJ parameters\cite{bian:2013} with MTP
  electrostatics,\cite{MM.eutectic:2024} blue), {\bf M2$^{\rm TIP3P}$}
  (red), and {\bf M2$^{\rm TIP4P}$} (green). Right panel: Hydration
  free energies $\Delta G_\mathrm{hyd}$ of a single SCN$^-$ ion in
  water solution. The dashed lines are experimental hydration free
  energies for N$_3^{-}$ and CN$^{-}$ and the reported $\Delta G_{\rm
    hyd} \sim -72$ kcal/mol for SCN$^{-}$.\cite{marcus:1997} For the
  density (left panel) the result for {\bf M2$^{\rm TIP4P}$} indicates
  that scaling $r_\mathrm{min}$ by 0.96 further improves the
  density. Evaluating the training set in Figure \ref{fig:models}B
  with this scaling leads to a deterioration of the overall RMSE by
  0.7\,kcal/mol, see Figure \ref{fig:models}C.}
\label{fig:rho_hfe}
\end{figure}

\noindent
The optimized energy functions {\bf M2$^{\rm TIP3P}$} and {\bf
  M2$^{\rm TIP4P}$} together with the previously
published\cite{MM.eutectic:2024} multipolar (MTP) electrostatic setup
{\bf M0} with literature LJ parameter\cite{bian:2013} were validated
in terms of the system density from $NpT$ simulation of a 3.8\,mol/kg
KSCN solution in water and the hydration free energy $\Delta
G_\mathrm{hyd}$ of a single SCN$^{-}$ in water. Figure
\ref{fig:rho_hfe} reports the results for differently scaled LJ
parameters $r_\mathrm{min}$ of the SCN$^-$ atoms by a factor $f$ in
the range from $f = 0.9$ to $f = 1.1$. The results are compared with
the measured density of an aqueous KSCN solution
($\rho_\mathrm{exp}=1.139$\,g/cm$^3$) at the equivalent KSCN molality
of $b(\mathrm{KSCN})=3.821$\,mol/kg\cite{albright:1992}, and the
estimated hydration free energy for SCN$^-$ of $\Delta G_\mathrm{hyd}
\sim -72$\,kcal/mol\cite{marcus:1997} which compares with related
anions such as HS$^-$ ($-74.0$\,kcal/mol), N$_3^-$
($-72.0$\,kcal/mol)\cite{pearson:1986}, or CN$^-$ ($-72.0 \pm
0.7$\,kcal/mol)\cite{pliego:2000}.\\

\noindent
Model {\bf M0} overestimates the experimental density and range of
$\Delta G_\mathrm{hyd}$ even with scaled LJ parameters
$r_\mathrm{min}$ by a factor of $f=1.1$, i.e. increased by 10\% in
order to reduce the electrostatic interactions.  Note that all results
from MD simulations using model {\bf M0} in the present work were
performed with scaled LJ parameter $r_\mathrm{min}$ by
$f_\mathbf{M0}=1.1$. On the other hand, the optimized energy function
{\bf M2$^{\rm TIP3P}$} shows good agreement even for
$f_\mathbf{M2,TIP3P}=1.0$ with an estimated density of $\rho =
1.119$\,g/cm$^3$. Energy function {\bf M2$^{\rm TIP4P}$} however,
shows the best match for a scaling factor $f_\mathbf{M2,TIP4P}=0.96$
with $\rho = 1.133$\,g/cm$^3$. In terms of the computed hydration free
energy of SCN$^{-}$, both optimized {\bf M2} energy functions yield
results within the expected range of the experiments (indicated by
dashed and dotted lines in Figure \ref{fig:rho_hfe}) for unscaled LJ
parameter ($f_\mathbf{M2,TIP3P} = f_\mathbf{M2,TIP4P} = 1.0$).  For
{\bf M2$^{\rm TIP4P}$} and a scaling factor of $f_\mathbf{M2,TIP4P} =
0.96$, where the density is matched best, $\Delta G_\mathrm{hyd}$ is
still within the expected range.\\

\noindent
The effect of the LJ parameter scaling $f_\mathbf{M2,TIP4P} = 0.96$
for the optimized energy functions {\bf M2$^{\rm TIP4P}$} on the
interaction energy correlation is shown in Figure
\ref{fig:models}C. The match between the fitted models $\Delta E_{\rm
  FF}$ and the {\it ab initio} values deteriorate somewhat with an
RMSE of $17.33$\,kcal/mol compared with the fully optimized LJ
parameter set $({\rm RMSE} = 16.61$\,kcal/mol).\\

\subsection{Structure and Ordering of the Mixture}
\begin{figure}
  \centering
  \includegraphics[width=0.95\textwidth]{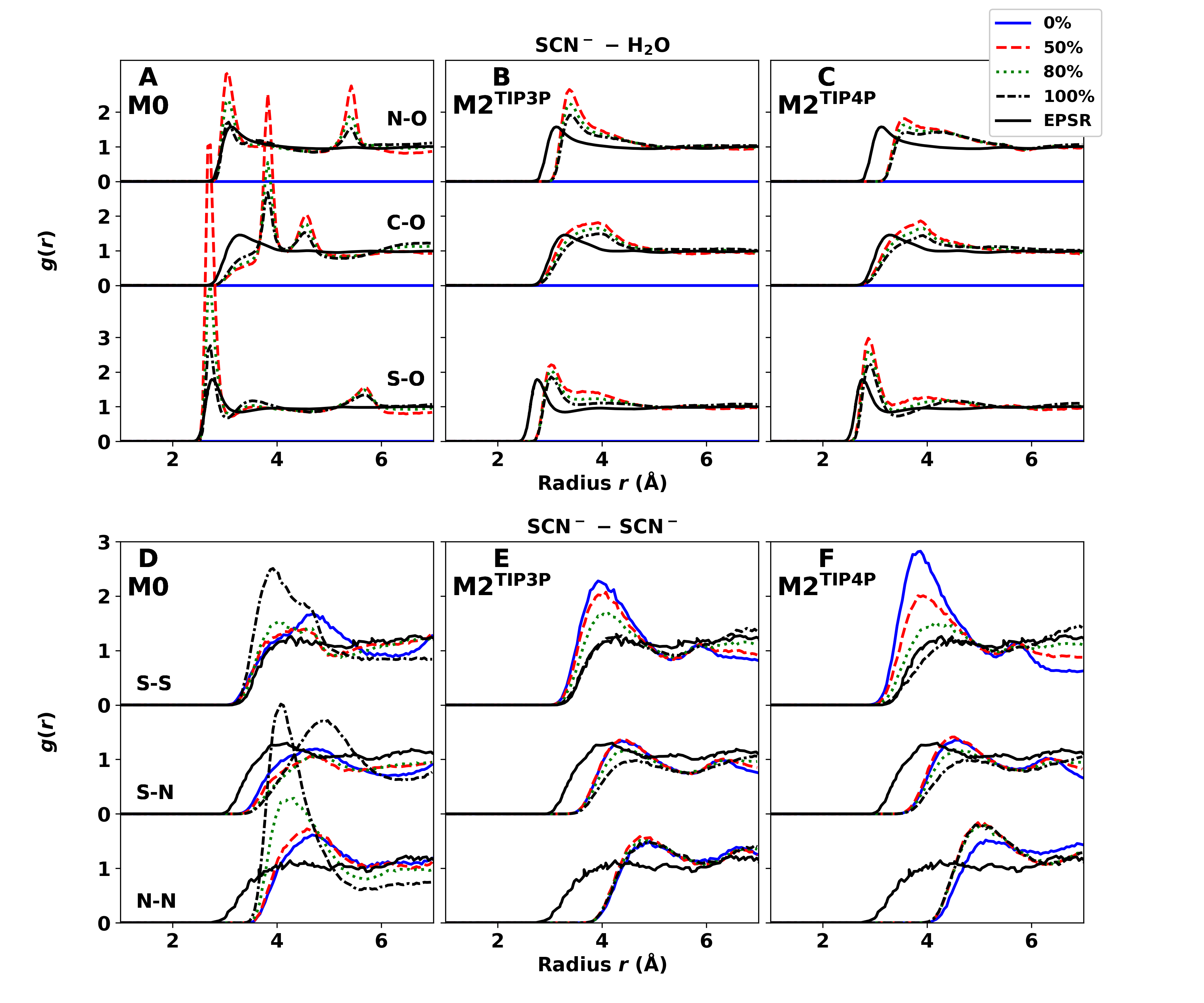}
\caption{Radial distribution function $g(r)$ between (A-C) SCN$^-$
  atoms and oxygen of water and (D-F) between both SCN$^-$ anions.
  The results are shown from simulations using (A, D) the MTP model
  and scaled LJ parameters, (B, E) the fMDCM approach and cluster
  fitted LJ parameters in TIP3P water and (C, F) the fMDCM approach
  and fitted LJ parameters in TIP4P water.  In comparison to the
  $3.8$\,mol/kg KSCN/water mixture (100\%), $g(r)$ of an aqueous KSCN
  solution ($2.5$\,mol/kg) obtained from empirical potential structure
  refinement (EPSR) to match experimentally measured neutron
  diffraction data are shown as solid black lines.\cite{botti:2009}}
\label{fig:g_fdcm}
\end{figure}

Next, MD simulations 25\,ns in length for the different compositions
considered were carried out using parametrization {\bf M2$^{\rm
    TIP4P}$}. From these simulations, radial pair distribution
functions $g(r)$ were determined and compared with results from
empirical potential structure refinement (EPSR) fits to best reproduce
neutron diffraction measurements of aqueous KSCN
solution.\cite{botti:2009} It is important to stress that such pair
distribution functions are not determined directly from experiments
but rather adjusted empirically to reproduce the neutron scattering
amplitude which is written as a weighted average over pair
distribution functions.\cite{botti:2009} Figure \ref{fig:g_fdcm}
reports SCN$^{-}$--water (top) and SCN$^{-}$--SCN$^{-}$ (bottom) pair
correlation functions for different mixtures KSCN in solution (0,
50\%, 80\%, 100\% water content) compared with data obtained from EPSR
(black).\\

\noindent
The radial distribution functions $g(r)$ in Figures \ref{fig:g_fdcm}B
and C (models {\bf M2$^{\rm TIP3P}$} and {\bf M2$^{\rm TIP4P}$}) are
considerably smoother than those using the MTP model in Figure
\ref{fig:g_fdcm}A.\cite{MM.eutectic:2024} For simulations in pure
water (100\%) direct comparisons between simulations (dot-dashed black
line) and EPSR results (solid black line) are
possible.\cite{botti:2009} For model {\bf M0}, the overall shape of
$g_{\rm N-O_{W}}(r)$ agrees reasonably well with the EPSR results
although the computed pair distribution function is overstructured
with a local maximum around 5.5\,\AA. Such maxima are not found for
models models {\bf M2$^{\rm TIP3P}$} and {\bf M2$^{\rm TIP4P}$} for
which the position of the maximum is shifted to larger separations and
the height of the maximum is overestimated for {\bf M2$^{\rm TIP3P}$}
but rather well described by {\bf M2$^{\rm TIP4P}$}. For both {\bf M2}
models the height of the first peak increases with decreasing water
content. In other words, with decreasing water density the ion
recruits water molecules. The shapes of $g(r_{\rm S/C/N-O})$ from
simulations using the two {\bf M2} models are consistent with those
inferred from measurements but typically the peak maxima are shifted
to longer separations $r$. A notable feature when using {\bf M2$^{\rm
    TIP4P}$} is the fact that the EPSR data shows some structure in
$g(r_{\rm S-O})$ around 3\,\AA\/ which is captured by {\bf M2$^{\rm
    TIP4P}$} but rather absent when using {\bf M2$^{\rm TIP3P}$}. On
the other hand, the maximum height of the first peak from $g(r_{\rm
  S-O})$ is consistent with the EPSR results for simulations with {\bf
  M2$^{\rm TIP3P}$} but clearly too pronounced for {\bf M2$^{\rm
    TIP4P}$}. Hence, interaction between SCN$^{-}$ and water is too
strong at the N-end of SCN$^{-}$ for {\bf M2$^{\rm TIP3P}$} but at the
S-end of SCN$^{-}$ for {\bf M2$^{\rm TIP4P}$}.\\

\noindent
The anion--anion pair distributions functions (Figure
\ref{fig:g_fdcm}) for the S--S contacts agree favourably between EPSR
results and simulations using M2$^{\rm TIP3P}$ and {\bf M2$^{\rm
    TIP4P}$}, see panels E and F. Using a MTP representation (Figure
\ref{fig:g_fdcm}D, model {\bf M0}) for the electrostatics leads to
overstructuring and the amplitude of the main peak is more than a
factor of 2 higher than that derived from the measurements. As the
water content decreases, the $g(r_{\rm SS})$ from {\bf M2$^{\rm
    TIP3P}$} and {\bf M2$^{\rm TIP4P}$} behave in a comparable
fashion. Only for the water-free system (blue) simulations using {\bf
  M2$^{\rm TIP4P}$} yield a higher maximum peak than from using {\bf
  M2$^{\rm TIP3P}$}.\\

\noindent
For the other two SCN$^{-}$--SCN$^{-}$ contacts considered ($g(r_{\rm
  SN})$ and $g(r_{\rm NN})$) the results from using M2$^{\rm TIP3P}$
and {\bf M2$^{\rm TIP4P}$} are comparable. The maxima of the first
peak are shifted to larger separations by $\sim 1$\,\AA\/ and the
maximum heights are somewhat under- and over-estimated,
respectively. Compared with simulations using multipolar model {\bf
  M0} using scaled literature LJ parameter by a scaling factor
$f_\mathbf{M0} = 1.1$, the {\bf M0} setup gets the same first peak
position for simulation in pure water (see dash-dotted black lines)
but the amplitudes are much too high.  The first peaks of $g(r_{\rm
  SN})$ are at larger distances then the EPSR data with a slightly
larger amplitude in the pure water simulation (Figure
\ref{fig:g_fdcm}D).\\

\noindent
The differences in the SCN$^{-}$--SCN$^{-}$ radial distribution
function from the simulation are a consequence of the different LJ
parameters, $\epsilon$ and $r_\mathrm{min}$, of the SCN$^{-}$ atoms
and their charges. Ignoring the electrostatic multipole components,
the MTP model in {\bf M0} assigns atomic charges of $q_\mathrm{MTP} =
\{-0.183, -0.362, -0.455\}$\,$e^-$ and the sum of distributed charges
to their nearest atoms SCN$^{-}$ in the fMDCM approach ({\bf M2$^{\rm
    TIP3P}$} and {\bf M2$^{\rm TIP4P}$}) are $q_\mathrm{fMDCM} =
\{-0.858, 1.000, -1.142\}$\,$e^-$ for the S, C and N atom,
respectively.  The large charge amplitudes in the electrostatic fMDCM
approach are the result of fitting the model ESP to the {\it ab
  initio} reference ESP, but it causes higher, generally more
attractive, electrostatic interaction contribution to the interaction
energies between SCN$^{-}$ and other residues such as water in
comparison to the smaller atomic charges in the MTP model.  Due to the
larger fMDCM charges, the fitted LJ parameters $r_\mathrm{min}$
parameter for SCN$^{-}$ atoms are generally larger (except for the
center carbon atom) than the scaled literature parameter in {\bf M0}
setup, see LJ parameters in Table \ref{sitab:params_m2tip3} and
\ref{sitab:params_m2tip4} in comparison to Table \ref{sitab:params}.
Larger LJ parameters $r_\mathrm{min}$ effect repulsive interaction
contribution earlier for decreasing nonbonding atom distances to
counter the more attractive electrostatic interaction, which also
affects the equilibrium SCN$^{-}$--SCN$^{-}$ pair distribution and
radial distribution functions in the solvent mixtures between the
models {\bf M0} and both {\bf M2$^{\rm TIP3P}$} and {\bf M2$^{\rm
TIP4P}$}.  \\

\begin{figure}
  \centering
  \includegraphics[width=0.80\textwidth]{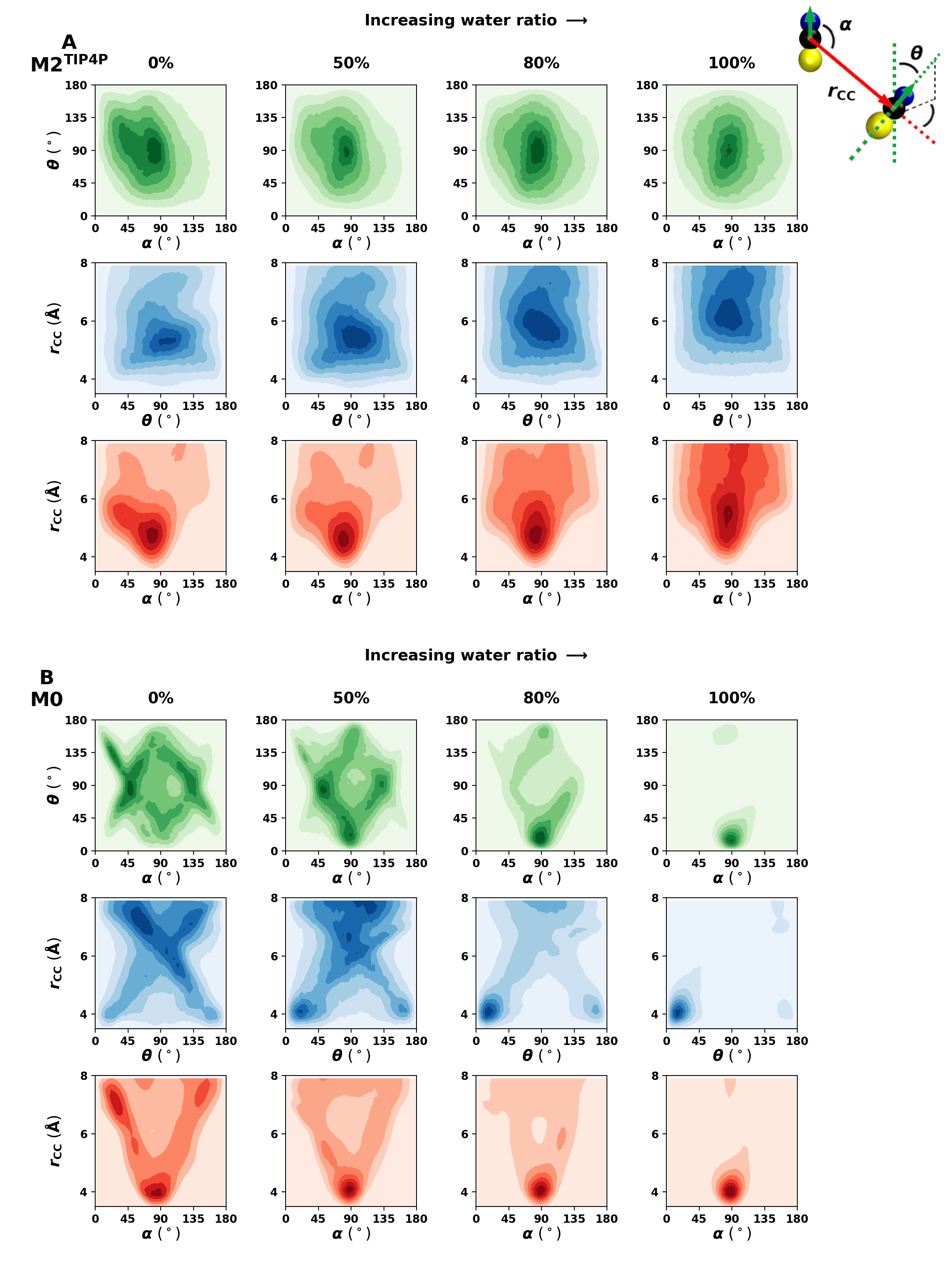}
\caption{Contour plots of the radial-angular distribution plots of
  SCN$^-$ pairs in different water/acetamide solutions (columns) from
  simulations using (A, {\bf M2$^{\rm TIP4P}$}) and {B, \bf M0}.  The
  data for model {\bf M0} is from previous work\cite{MM.eutectic:2024}
  and the coordinate system as in the upper right-hand side.}
\label{fig:radang2}
\end{figure}

\noindent
Two-dimensional distributions functions $P(\alpha,r_{\rm CC})$,
$P(\theta,r_{\rm CC})$, and $P(\alpha,\theta)$ in Figure
\ref{fig:radang2} provide a complementary view of the relative
orientation of the anions. The definition of the coordinates is also
provided in this Figure. The bottom row ($P(\alpha,r_{\rm CC})$) in
Figure \ref{fig:radang2}A shows that with increasing water content the
average C--C separation in simulations using the {\bf M2$^{\rm
    TIP4P}$} model shifts to larger values. This is consistent with
Figure \ref{fig:g_fdcm}F and with simulations using {\bf M2$^{\rm
    TIP3P}$}, see Figure \ref{fig:g_fdcm}D.  On the other hand, the
maximum of the relative angular orientation, described by $\alpha$,
remains the same for all solvent mixtures although the angular
constraint tightens as the water concentration increases. Finally, the
average azimuthal orientation $\theta$ remains around $90^\circ$ for
all solvent compositions. These distributions contrast with those from
simulations using model {\bf M0} which are reported in Figure
\ref{fig:radang2}B.  Atomic multipoles have pronounced directionality
as they are based on p- and d-orbitals if moments up to quadrupole are
included. This directionality can lead to overstructuring which is
particularly prevalent for the $\theta-$direction (green and blue
distributions).\\

\subsection{Dynamics of the Mixtures}
\begin{figure}
\centering \includegraphics[width=0.90\textwidth]{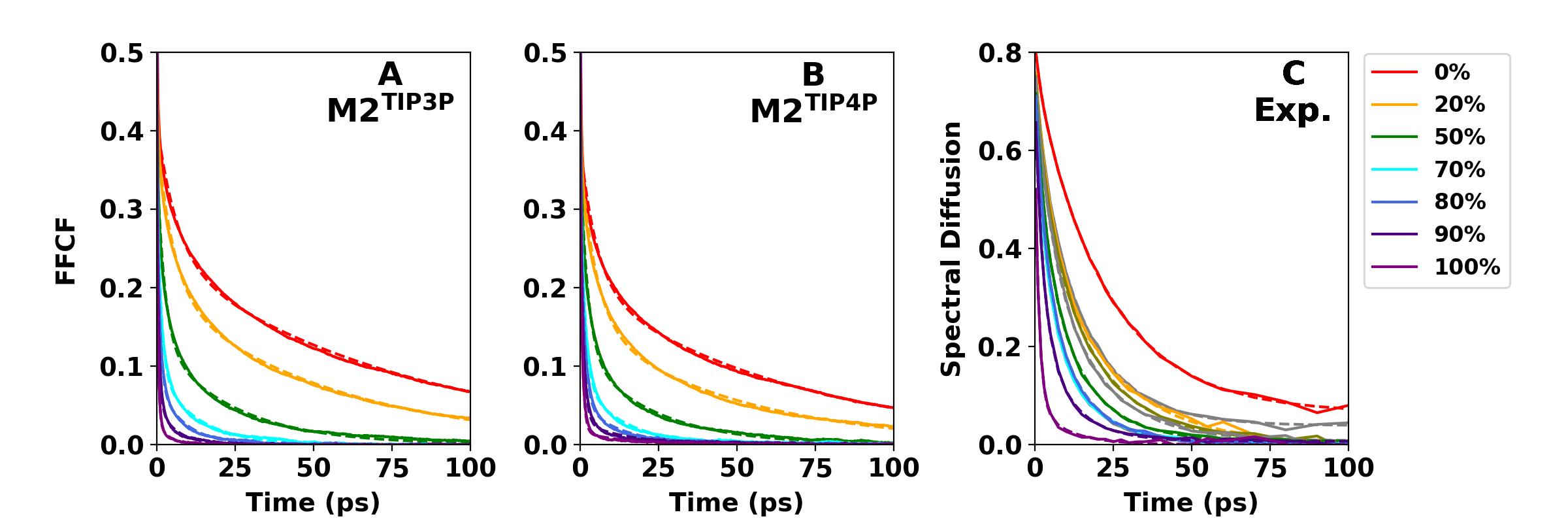}
\caption{Panels A and B: FFCF (solid lines) of the INM frequencies
  $\nu_3$ of SCN$^-$ anion from simulation with water models for
  different mixing ratios of acetamide and water. Panel C: Measured
  spectral diffusion.\cite{MM.eutectic:2024} Gray lines in panel C are
  experimental results for which no simulations were carried out. A
  bi-exponential function (dashed lines) was fit to each computed FFCF
  (A, B) and to the experimental data in panel C in the range from
  $0.25$\,ps to 1\,ns. All computed FFCFs are normalized.}
\label{fig:ffcf}
\end{figure}

Next, the dynamics of the mixtures was studied by analyzing the C--N
vibrations from instantaneous normal modes\cite{stratt:1994} of the
SCN$^-$ anion. For this, the frequency fluctuation correlation
functions (FFCF) for varying water content were determined, see Figure
\ref{fig:ffcf}A/B for the FFCFs from simulations with models {\bf
  M2$^{\rm TIP3P}$} and {\bf M2$^{\rm TIP4P}$}, respectively. Figure
\ref{fig:ffcf}C reports the experimentally measured FFCFs from the
2D-IR experiments.\cite{MM.eutectic:2024} The FFCFs feature a
pronounced dependence on the water content from 0\% (red) to 100\%
(purple) water fraction and the computed decay curves rather
realistically describe the measurements for both optimized models.
Assuming a single exponential function for representing the computed
FFCFs leads to a considerably deteriorated fit compared with assuming
two time scales. Hence, for the FFCFs from the simulations a
bi-exponential fit was used as a meaningful model for all mixtures
from which two time scales $\tau_{\rm fast}$ and $\tau_{\rm slow}$
were extracted.\\

\begin{figure}
\centering \includegraphics[width=0.90\textwidth]{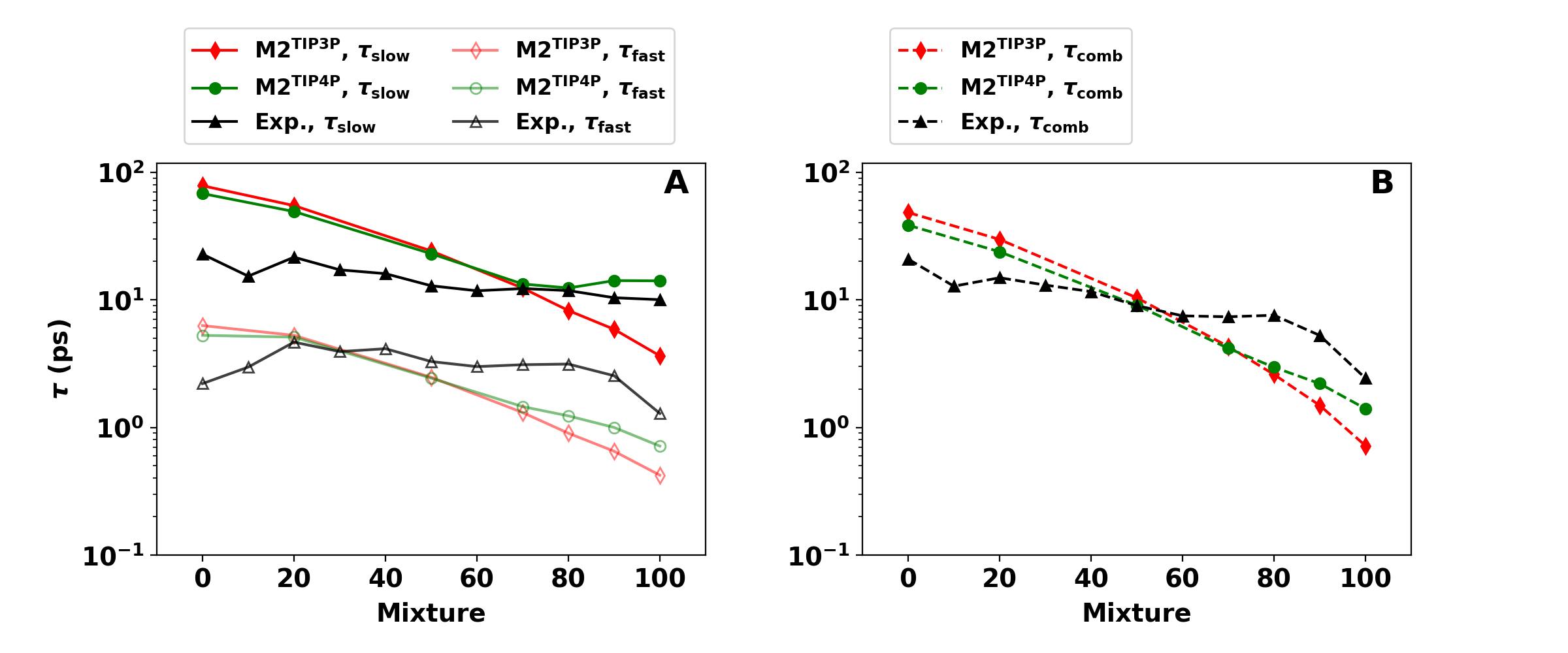}
\caption{Panel A: Fitted lifetimes (log-scale) $\tau_\mathrm{slow}$
  (full marker) and $\tau_\mathrm{fast}$ (open marker) of a
  bi-exponential function to the FFCF of the INM frequencies $\nu_3$
  of the SCN$^-$ anion from simulation with different water
  models. The lifetimes from the bi-exponential fit to the
  experimental spectral diffusion are shown in black. Panel B:
  Amplitude-weighted decay times $\tau_{\rm comb}$ for all results,
  see text. For results on a linear scale, see Figure
  \ref{sifig:ffcf_fit}.}
\label{fig:times}
\end{figure}

\noindent
To allow direct comparison between experiment and simulations, the
experimental data was also fit to a double-exponential, see
Eq. \ref{eq:ffcffit}A. The resulting lifetimes $\tau_{\rm fast}$ and
$\tau_{\rm slow}$ are shown in Figure \ref{fig:times}. All optimized
parameters of the bi-exponential fit are reported in Figure
\ref{sifig:ffcf_fit} and Table \ref{sitab:ffcf_fit}.  Qualitatively,
the two {\bf M2} models capture the experimentally observed FFCFs and
their dependence on the amount of water in the solvent mixture.\\

\noindent
A quantitative analysis and comparison with measurements is afforded
by considering the decay times of the FFCFs depending on water
content, see Figure \ref{fig:times}. Experimentally, the slow decay
$\tau_{\rm slow}^{\rm exp}$ (black filled triangles) decreases almost
monotonously with increasing water content. This is qualitatively
reproduced from simulations using both models. However, it is
interesting to note that for the highest water content the {\bf
  M2$^{\rm TIP4P}$} model is rather consistent with the experiment
whereas for {\bf M2$^{\rm TIP3P}$} $\tau_{\rm slow}$ continues to
decrease with increasing water content. For the fast decay $\tau_{\rm
  fast}^{\rm exp}$ (black open triangles) an increase with increasing
water content up to $\sim 30$ \% is observed followed by a monotonous
decrease. Both, {\bf M2$^{\rm TIP3P}$} and {\bf M2$^{\rm TIP4P}$}, do
not reproduce the plateau between 20 \% and 40 \%. On the other hand,
the decrease for $\tau_{\rm fast}$ above 60 \% water content is
realistically described. From a quantitative perspective, results from
simulations using {\bf M2$^{\rm TIP4P}$} are in somewhat closer
agreement with experiment.\\

\noindent
It is, however, known that in biexponential fits the parameters can be
strongly correlated. Therefore, amplitude-weighted decay times
$\tau_{\rm comb} = a_{\rm slow} \tau_{\rm slow} + a_{\rm fast}
\tau_{\rm fast}$ were also considered, see Figure \ref{fig:times}B,
with amplitudes reported in Figure \ref{sifig:ffcf_fit}A. Again, the
computations reproduce the general trend: the decrease in $\tau_{\rm
  comb}$ with increasing water content. However, simulations using
both {\bf M2} models considered here feature a somewhat too steep
decrease with increasing water content except for {\bf M2$^{\rm
    TIP4P}$} for the highest water ratio.\\

\begin{figure}
\centering
\includegraphics[width=0.50\textwidth]{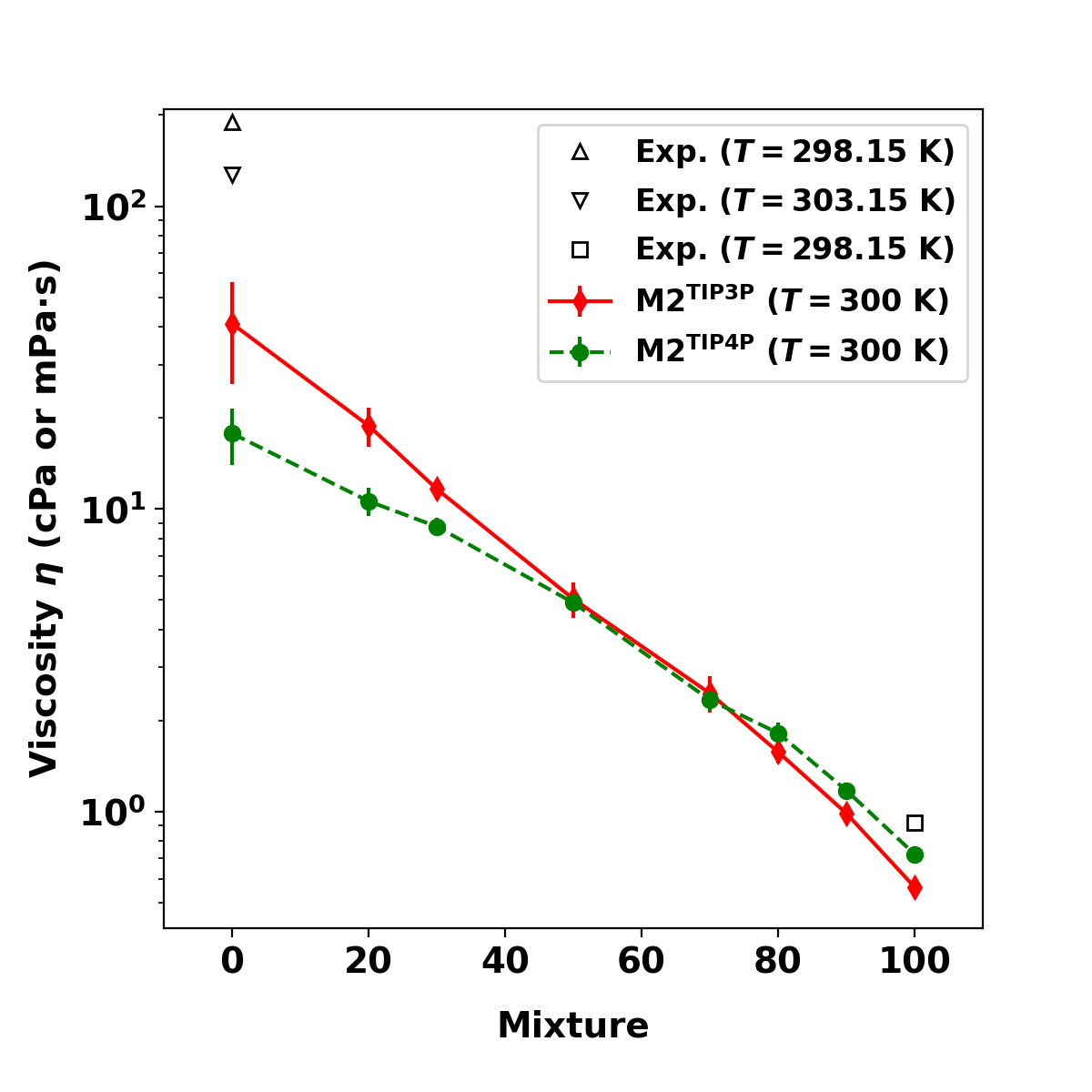}
\caption{Computed viscosity from the Green-Kubo relation (see main
  text) using the stress tensor correlation function from simulations
  of KSCN in acetamide/TIP3P {\bf M2$^{\rm TIP3P}$} and
  acetamide/TIP4P mixtures {\bf M2$^{\rm TIP4P}$}. The results were
  obtained from 5 individual runs of 5\,ns $NVT$ simulations each per
  mixture. The standard deviation of the viscosity estimations per run
  are shown as error bar. It is important to stress that the
  simulation length (5\,ns) is not sufficient for convergence, in
  particular for low water content.  Experimental measured viscosities
  for KSCN in acetamide (upper and lower triangle)\cite{liu:2016} and
  in water (open square)\cite{albright:1992} are shown but at
  different temperatures.}
\label{fig:viscosity}
\end{figure}

\noindent
Finally, simulations of KSCN in acetamide using models {\bf M2$^{\rm
    TIP3P}$} and {\bf M2$^{\rm TIP4P}$} yield average viscosities of
$40.9 \pm 15.0$\,mPa$\cdot$s and $17.8\pm3.8$\,mPa$\cdot$s,
respectively, that differ by a factor of 3 for the water-free system
(i.e. only acetamide as the solvent) which underestimate the
experimentally reported value $\eta = 127.3$\,mPa$\cdot$s at
303.15\,K.\cite{liu:2013,liu:2016} With increasing water content
$\eta$ decreases exponentially which is correctly captured from both
{\bf M2} models. For KSCN in pure water, the experimentally reported
viscosity of $0.92$\,mPa$\cdot$s (for
$c=3.355$\,mol/l)\cite{albright:1992} at lower temperature of
298.15\,K compares with $0.56$\,mPa$\cdot$s ({\bf M2$^{\rm TIP3P}$})
and $0.72$\,mPa$\cdot$s ({\bf M2$^{\rm TIP4P}$}) at 300\,K, which is
rather encouraging. Hence, part of the disagreement for low-water
content can probably be mitigated by including acetamide in the fitting
procedure. With respect to comparing results from simulations and
experiments it is worthwhile to mention that the experiments hint
towards a pronounced effect of temperature on $\eta$: increasing $T$
by 10 K reduces $\eta$ by a factor of two. It should also be noted
that further converging $\eta$ requires probably longer simulation
times. From a computer efficiency perspective it appears to be
advisable to consider $\eta$ as a validation property rather than a
property used for fitting the intermolecular interactions.\\

\noindent
The smaller viscosities of the KSCN in different acetamide/water
mixtures using both {\bf M2} models can be related to the larger LJ
parameter $r_\mathrm{min}$ than in literature for SCN$^-$. Even though
the LJ parameter is optimized to fit reference interaction energies
between SCN$^-$ and water-containing cluster shells, no fits to
interaction energies including acetamide have been done. In general,
increased LJ parameters $r_\mathrm{min}$ on SCN$^-$ increase space
between the anion and other surrounding species (see radial
distribution functions in Figure \ref{fig:g_fdcm}) which leads to
weaker interactions with acetamide, increased mobility and hence
reduced viscosity.\\

\section{Conclusions}
The present work validates a cluster-based parametrization workflow
for energy functions of electrostatically driven systems (here
eutectic liquids) based on quantum chemical interaction
energies. Irrespective of the water model used (TIP3P or TIP4P) the
thermodynamic (density $\rho$, hydration free energy $\Delta
G_\mathrm{hyd}$), structural (pair distribution functions $g(r)$),
spectroscopic (2d-infrared), and transport (viscosity $\eta$)
properties are realistically described by using models {\bf M2$^{\rm
    TIP3P}$} and {\bf M2$^{\rm TIP4P}$} from extensive MD
simulations.\\

\noindent
The fact that the cluster-based approach is largely insensitive to the
water model used (here TIP3P vs. TIP4P) indicates that other, yet
``better'' water models (BWM) can be employed to conceive models {\bf
  M2$^{\rm BWM}$}. The performance of such models is expected to be
particularly realistic for high-water content whereas for low-water
content the parametrization protocol needs to include the cosolvent
(here acetamide).\\

\noindent
In conclusion, the present work demonstrates that in the absence or
limited availability of reference measurements to validate such
parametrizations, conceiving energy functions following model {\bf M2}
provides a meaningful starting point for (semi-)quantitative
simulations, depending on the property considered. It will be
interesting to apply the present approach to other compositions and
different systems, such as ionic liquids.

\section*{Acknowledgment}
This work has been financially supported by the Swiss National Science
Foundation (NCCR MUST, 200020\_219779, 200021\_215088 to MM), the
University of Basel (to MM) by European Union's Horizon 2020 research
and innovation program under the Marie Sk{\l}odowska-Curie grant
agreement No 801459 -FP-RESOMUS (to KT). \\

\section*{Supporting Information}
The supporting material includes Table \ref{sitab:composition} to
\ref{sitab:ffcf_fit} and Figure \ref{sifig:ffcf_fit}.

\section*{Data Availability}
Relevant data for the present study are available at
\url{https://github.com/MMunibas/DES3}.

\bibliography{refs}

\clearpage

\renewcommand{\thepage}{S\arabic{page}}
\renewcommand{\thetable}{S\arabic{table}}
\renewcommand{\thefigure}{S\arabic{figure}}
\renewcommand{\theequation}{S\arabic{equation}}
\renewcommand{\thesection}{S\arabic{section}} 

\newpage

\noindent
{\bf Supporting Information: Force Fields for Deep Eutectic Systems:
  K$^+$/SCN$^-$ in Water/Acetamide Mixtures}

\section{Supporting Figures and Tables}

\begin{table}
\caption{Molar fraction and number of molecules (Water/W,
  Acetamide/ACM, K$^+$ and SCN$^{-}$) used on the MD simulations for
  each component of the systems for given W:ACM mixing
  ratio.}
\label{sitab:composition}
\begin{tabular}{c||ccc||ccc}
\hline\hline
& \multicolumn{3}{c||}{Molar Fraction} & \multicolumn{3}{c}{Number of Molecules} \\ \hline
W:ACM & \multicolumn{1}{c|}{H$_{2}$O}   & \multicolumn{1}{c|}{Acetamide} & KSCN  & \multicolumn{1}{c|}{H$_{2}$O} & \multicolumn{1}{c|}{Acetamide} & KSCN \\
\hline \hline
100                            & \multicolumn{1}{c|}{0.901} & \multicolumn{1}{c|}{0.000}     & 0.099 & \multicolumn{1}{c|}{685} & \multicolumn{1}{c|}{0}         & 75   \\
90                             & \multicolumn{1}{c|}{0.791} & \multicolumn{1}{c|}{0.092}     & 0.118 & \multicolumn{1}{c|}{506} & \multicolumn{1}{c|}{59}        & 75   \\
80                             & \multicolumn{1}{c|}{0.686} & \multicolumn{1}{c|}{0.179}     & 0.135 & \multicolumn{1}{c|}{381} & \multicolumn{1}{c|}{100}       & 75   \\
70                             & \multicolumn{1}{c|}{0.586} & \multicolumn{1}{c|}{0.263}     & 0.152 & \multicolumn{1}{c|}{290} & \multicolumn{1}{c|}{130}       & 75   \\
50                             & \multicolumn{1}{c|}{0.399} & \multicolumn{1}{c|}{0.418}     & 0.183 & \multicolumn{1}{c|}{164} & \multicolumn{1}{c|}{171}       & 75   \\
40                             & \multicolumn{1}{c|}{0.312} & \multicolumn{1}{c|}{0.490}     & 0.198 & \multicolumn{1}{c|}{119} & \multicolumn{1}{c|}{186}       & 75   \\
30                             & \multicolumn{1}{c|}{0.229} & \multicolumn{1}{c|}{0.559}     & 0.212 & \multicolumn{1}{c|}{81}  & \multicolumn{1}{c|}{198}       & 75   \\
20                             & \multicolumn{1}{c|}{0.149} & \multicolumn{1}{c|}{0.626}     & 0.225 & \multicolumn{1}{c|}{50}  & \multicolumn{1}{c|}{209}       & 75   \\
0                              & \multicolumn{1}{c|}{0.000} & \multicolumn{1}{c|}{0.750}     & 0.250 & \multicolumn{1}{c|}{0}   & \multicolumn{1}{c|}{225}       & 75   \\
\hline
\hline
\end{tabular}
\end{table}

\begin{table}
\caption{Bonded and non-bonded parameters for Model {\bf M0} using
  atomic multipoles (MTP) up to quadrupole.  Nonbonded LJ parameters
  for SCN$^-$ are adopted from Ref. \citenum{bian:2013} with
  $r_\mathrm{min}$ scaled by $f=1.1$.}
\label{sitab:params}
\begin{tabular}{c|ccc}
\hline\hline
\textbf{Residues} & \multicolumn{3}{c}{Parameters} \\
\hline \hline
\textbf{Acetamide} & \multicolumn{3}{c}{CGenFF\cite{cgenff}} \\
\hline \hline
\textbf{Water} & \multicolumn{3}{c}{TIP3P\cite{TIP3P-Jorgensen-1983}} \\
\hline \hline
\textbf{K$^+$} & \multicolumn{3}{c}{} \\
\hline
Nonbonded\cite{bian:2013} & $\epsilon$ (kcal/mol)& $r_\mathrm{min}$ (\AA) & $q$ (e)\\
K$^+$ & $0.1004$ & $3.7378$ & $+1$ \\
\hline \hline 
\textbf{SCN$^-$} & \multicolumn{3}{c}{} \\
\hline
Bonded & \multicolumn{3}{c}{RKHS (see $^a$)} \\
\hline 
Nonbonded & $\epsilon$ (kcal/mol)& $r_\mathrm{min}$ (\AA) & $q$ (e) \\
S & $0.3639$ & $4.3462\,(=1.1\cdot3.9510)$ & $-0.183$ \\
C & $0.0741$ & $4.0870\,(=1.1\cdot3.7154)$ & $-0.455$ \\
N & $0.1016$ & $4.1362\,(=1.1\cdot3.7602)$ & $-0.362$ \\
\hline 
Atomic Multipoles & S & C & N \\
$Q_{00}$  & $-0.183$ & $-0.362$ & $-0.455$ \\
$Q_{10}$  & $1.179$ & $0.163$ & $0.319$ \\
$Q_{11c}$ & $0.0$ & $0.0$ & $0.0$ \\
$Q_{11s}$ & $0.0$ & $0.0$ & $0.0$ \\
$Q_{20}$  & $-1.310$ & $-0.929$ & $-3.114$ \\
$Q_{21c}$ & $0.0$ & $0.0$ & $0.0$ \\
$Q_{21s}$ & $0.0$ & $0.0$ & $0.0$ \\
$Q_{22c}$ & $0.0$ & $0.0$ & $0.0$ \\
$Q_{22s}$ & $0.0$ & $0.0$ & $0.0$ \\
\hline \hline
\end{tabular}
$^a$\url{https://github.com/MMunibas/DES2/blob/main/M0/source/rkhs_SCN_rRz.csv}
\end{table}

\begin{table}
\caption{Bonded and non-bonded parameters for simulation using the
  TIP3P water model labeled as {\bf M2$^{\rm TIP3P}$}. Nonbonded LJ
  parameters for SCN$^-$ are adopted from cluster interaction energy
  fit.}
\label{sitab:params_m2tip3}
\begin{tabular}{c|ccc}
\hline\hline
\textbf{Residues} & \multicolumn{3}{c}{Parameters} \\
\hline \hline
\textbf{Acetamide} & \multicolumn{3}{c}{CGenFF\cite{cgenff}} \\
\hline \hline
\textbf{Water} & \multicolumn{3}{c}{TIP3P\cite{TIP3P-Jorgensen-1983}} \\
\hline \hline
\textbf{K$^+$} & \multicolumn{3}{c}{} \\
\hline
Nonbonded\cite{bian:2013} & $\epsilon$ (kcal/mol)& $r_\mathrm{min}$ (\AA) & $q$ (e)\\
K$^+$ & $0.1004$ & $3.7378$ & $+1$ \\
\hline \hline 
\textbf{SCN$^-$} & \multicolumn{3}{c}{} \\
\hline
Bonded & \multicolumn{3}{c}{RKHS (see $^a$)} \\
\hline 
Nonbonded & $\epsilon$ (kcal/mol)& $r_\mathrm{min}$ (\AA) & $q$ (e) \\
S & $0.1836$ & $4.8558$ & $-$ \\
C & $0.0001$ & $3.7868$ & $-$ \\
N & $0.0223$ & $4.5720$ & $-$ \\
\hline 
Electrostatic Model &  \multicolumn{3}{c}{fMDCM (see $^b$)}\\
\hline \hline
\end{tabular}
$^a$\url{https://github.com/MMunibas/DES2/blob/main/M2/source/rkhs_SCN_rRz.csv}
$^b$\url{https://github.com/MMunibas/DES2/blob/main/M2/source/scn_fluc.dcm}
\end{table}

\begin{table}
\caption{Bonded and non-bonded parameters simulation using the TIP4P
  water model labeled as {\bf M2$^{\rm TIP4P}$}.  Nonbonded LJ
  parameters for SCN$^-$ are adopted from cluster interaction energy
  fit.}
\label{sitab:params_m2tip4}
\begin{tabular}{c|ccc}
\hline\hline
\textbf{Residues} & \multicolumn{3}{c}{Parameters} \\
\hline \hline
\textbf{Acetamide} & \multicolumn{3}{c}{CGenFF\cite{cgenff}} \\
\hline \hline
\textbf{Water} & \multicolumn{3}{c}{TIP4P\cite{TIP3P-Jorgensen-1983}} \\
\hline \hline
\textbf{K$^+$} & \multicolumn{3}{c}{} \\
\hline
Nonbonded\cite{bian:2013} & $\epsilon$ (kcal/mol)& $r_\mathrm{min}$ (\AA) & $q$ (e)\\
K$^+$ & $0.1004$ & $3.7378$ & $+1$ \\
\hline \hline 
\textbf{SCN$^-$} & \multicolumn{3}{c}{} \\
\hline
Bonded & \multicolumn{3}{c}{RKHS (see $^a$)} \\
\hline 
Nonbonded & $\epsilon$ (kcal/mol)& $r_\mathrm{min}$ (\AA) & $q$ (e) \\
S & $0.1118$ & $5.3984$ & $-$ \\
C & $0.0001$ & $3.1786$ & $-$ \\
N & $0.0037$ & $5.0388$ & $-$ \\
\hline 
Electrostatic Model &  \multicolumn{3}{c}{fMDCM (see $^b$)}\\
\hline \hline
\end{tabular}
$^a$\url{https://github.com/MMunibas/DES2/blob/main/M3/source/rkhs_SCN_rRz.csv}
$^b$\url{https://github.com/MMunibas/DES2/blob/main/M3/source/scn_fluc.dcm}
\end{table}

\begin{figure}
\centering
\includegraphics[width=0.90\textwidth]{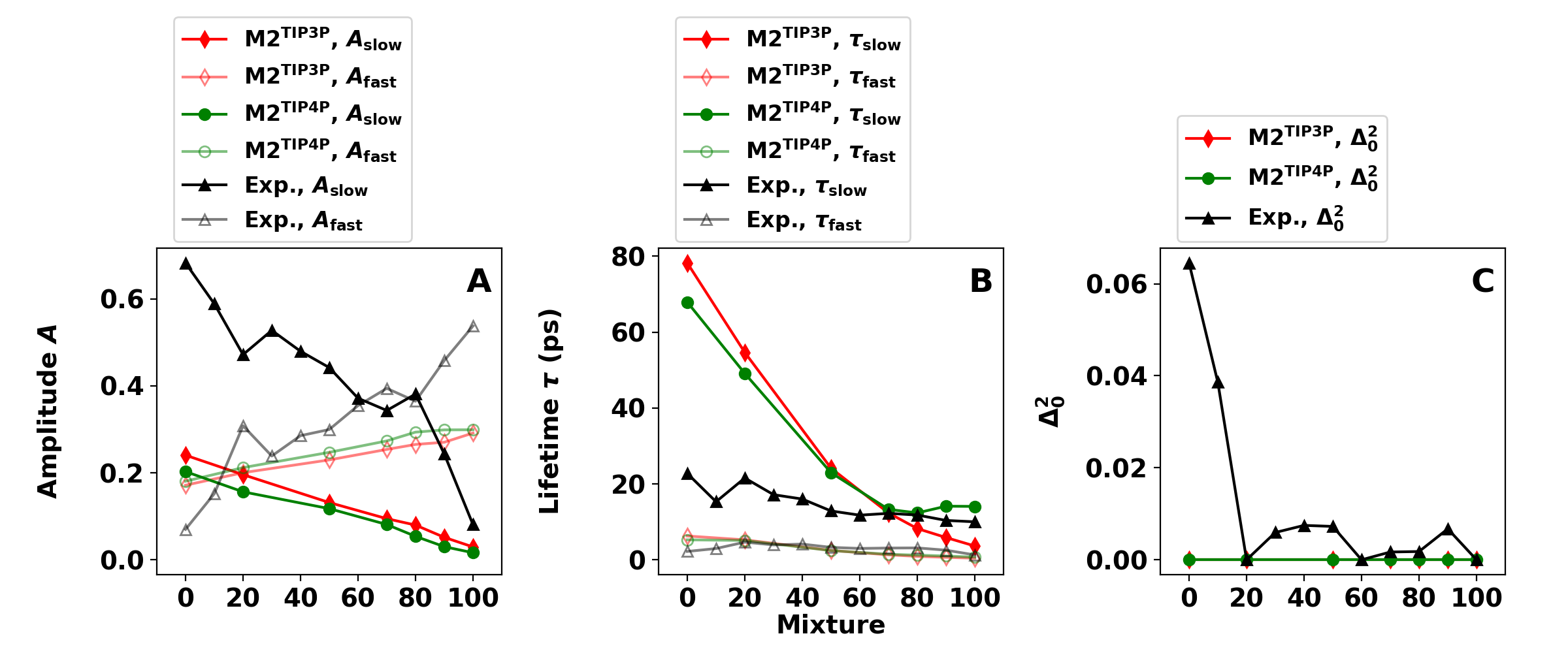}
\caption{Panels A to C: Optimized amplitudes, $(\tau_{\rm
    fast},\tau_{\rm slow})$ (linear scale) and $\Delta_0^2$ of the
  bi-exponential fits to the computed frequency-frequency correlation
  functions and the spectral diffusion. The vibrational mode
  considered was $\nu_3$ of SCN$^-$ and frequencies were determined
  from instantaneous normal modes.}
\label{sifig:ffcf_fit}
\end{figure}

\begin{table}
\caption{Fitted lifetimes in ps to the ({\bf M0, M1, M2}) computed
  FFCFs and ({\bf Exp.})  experimentally measured spectral diffusion
  from KSCN in water(W)/acetamide(ACM) mixtures for given W:ACM mixing
  ratios.}
\label{sitab:ffcf_fit}
\begin{tabular}{c||cc||cc||cc}
\hline\hline
  & \multicolumn{2}{c||}{{\bf M2$^{\rm TIP3P}$}} & \multicolumn{2}{c||}{{\bf M2$^{\rm TIP4P}$}} & \multicolumn{2}{c}{\bf Exp.} \\ \hline
W:ACM & 
  \multicolumn{1}{c}{$\tau_{\rm fast}$}   & \multicolumn{1}{c||}{$\tau_{\rm slow}$} &
  \multicolumn{1}{c}{$\tau_{\rm fast}$}   & \multicolumn{1}{c||}{$\tau_{\rm slow}$} &
  \multicolumn{1}{c}{$\tau_{\rm fast}$}   & \multicolumn{1}{c}{$\tau_{\rm slow}$} \\
\hline \hline
0 & 6.28 & 78.21 & 5.26 & 67.89 & 2.20 & 22.74 \\
10 & - & - & - & - & 2.97 & 15.30 \\
20 & 5.26 & 54.64 & 5.10 & 49.11 & 4.66 & 21.55 \\
30 & - & - & - & - & 3.92 & 17.16 \\
40 & - & - & - & - & 4.14 & 16.01 \\
50 & 2.47 & 24.14 & 2.43 & 22.95 & 3.28 & 12.87 \\
60 & - & - & - & - & 2.99 & 11.76 \\
70 & 1.31 & 12.33 & 1.46 & 13.29 & 3.10 & 12.24 \\
80 & 0.90 & 8.23 & 1.23 & 12.36 & 3.14 & 11.80 \\
90 & 0.65 & 5.87 & 1.00 & 14.13 & 2.54 & 10.37 \\
100 & 0.42 & 3.64 & 0.71 & 14.07 & 1.29 & 10.01 \\
\hline
\hline
\end{tabular}
\end{table}

\clearpage

\end{document}